\documentclass[twoside,twocolumn,9pt]{article}
\usepackage{extsizes}
\usepackage{dblfloatfix}
\usepackage[super,sort&compress,comma]{natbib} 
\usepackage[version=3]{mhchem}
\usepackage[left=1.5cm, right=1.5cm, top=1.785cm, bottom=2.0cm]{geometry}
\usepackage{balance}
\usepackage{mathptmx}
\usepackage{sectsty}
\usepackage{graphicx} 
\usepackage{lastpage}
\usepackage[format=plain,justification=justified,singlelinecheck=false,font={stretch=1.125,small,sf},labelfont=bf,labelsep=space]{caption}
\usepackage{float}
\usepackage{fancyhdr}
\usepackage{fnpos}
\usepackage[english]{babel}
\addto{\captionsenglish}{%
  
}
\usepackage{array}
\usepackage{droidsans}
\usepackage{charter}
\usepackage[T1]{fontenc}
\usepackage[usenames,dvipsnames]{xcolor}
\usepackage{setspace}
\usepackage[compact]{titlesec}
\usepackage{hyperref}
\usepackage{pdfpages}
\usepackage[compatibility=false]{subcaption}

\usepackage{soul}

\usepackage{epstopdf}

\definecolor{cream}{RGB}{222,217,201}
\begin{document}

\pagestyle{fancy}
\thispagestyle{plain}
\fancypagestyle{plain}{
\renewcommand{\headrulewidth}{0pt}
}

\makeFNbottom
\makeatletter
\renewcommand\LARGE{\@setfontsize\LARGE{15pt}{17}}
\renewcommand\Large{\@setfontsize\Large{12pt}{14}}
\renewcommand\large{\@setfontsize\large{10pt}{12}}
\renewcommand\footnotesize{\@setfontsize\footnotesize{7pt}{10}}
\makeatother

\renewcommand{\thefootnote}{\fnsymbol{footnote}}
\renewcommand\footnoterule{\vspace*{1pt}%
\color{cream}\hrule width 3.5in height 0.4pt \color{black}\vspace*{5pt}} 
\setcounter{secnumdepth}{5}

\makeatletter 
\renewcommand\@biblabel[1]{#1}            
\renewcommand\@makefntext[1]%
{\noindent\makebox[0pt][r]{\@thefnmark\,}#1}
\makeatother 
\renewcommand{\figurename}{\small{Fig.}~}
\sectionfont{\sffamily\Large}
\subsectionfont{\normalsize}
\subsubsectionfont{\bf}
\setstretch{1.125} 
\setlength{\skip\footins}{0.8cm}
\setlength{\footnotesep}{0.25cm}
\setlength{\jot}{10pt}
\titlespacing*{\section}{0pt}{4pt}{4pt}
\titlespacing*{\subsection}{0pt}{15pt}{1pt}

\fancyfoot{}
\fancyfoot[LO,RE]{\vspace{-7.1pt}\includegraphics[height=9pt]{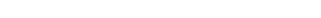}}
\fancyfoot[CO]{\vspace{-7.1pt}\hspace{13.2cm}\includegraphics{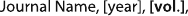}}
\fancyfoot[CE]{\vspace{-7.2pt}\hspace{-14.2cm}\includegraphics{head_foot/RF}}
\fancyfoot[RO]{\footnotesize{\sffamily{1--\pageref{LastPage} ~\textbar  \hspace{2pt}\thepage}}}
\fancyfoot[LE]{\footnotesize{\sffamily{\thepage~\textbar\hspace{3.45cm} 1--\pageref{LastPage}}}}
\fancyhead{}
\renewcommand{\headrulewidth}{0pt} 
\renewcommand{\footrulewidth}{0pt}
\setlength{\arrayrulewidth}{1pt}
\setlength{\columnsep}{6.5mm}
\setlength\bibsep{1pt}

\makeatletter 
\newlength{\figrulesep} 
\setlength{\figrulesep}{0.5\textfloatsep} 

\newcommand{\topfigrule}{\vspace*{-1pt}%
\noindent{\color{cream}\rule[-\figrulesep]{\columnwidth}{1.5pt}} }

\newcommand{\botfigrule}{\vspace*{-2pt}%
\noindent{\color{cream}\rule[\figrulesep]{\columnwidth}{1.5pt}} }

\newcommand{\dblfigrule}{\vspace*{-1pt}%
\noindent{\color{cream}\rule[-\figrulesep]{\textwidth}{1.5pt}} }

\makeatother

\twocolumn[
  \begin{@twocolumnfalse}
{\includegraphics[height=30pt]{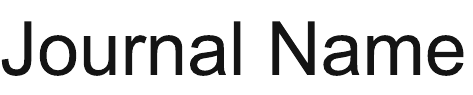}\hfill\raisebox{0pt}[0pt][0pt]{\includegraphics[height=55pt]{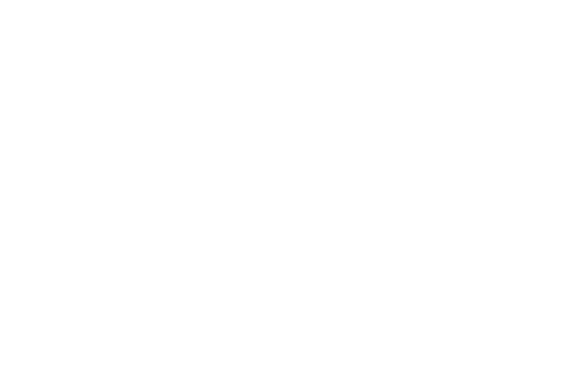}}\\[1ex]
\includegraphics[width=18.5cm]{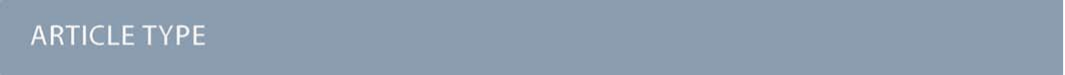}}\par
\vspace{1em}
\sffamily
\begin{tabular}{m{4.5cm} p{13.5cm} }

\includegraphics{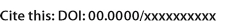} & \noindent\LARGE{\textbf{Beyond ideal cavities: quantifying the impact of cavity dissipation
on light–matter strong coupling$^\dag$}} \\
\vspace{0.3cm} & \vspace{0.3cm} \\

 & \noindent\large{Giovanna Bruno,$^{\ast}$\textit{$^{a}$} Enrico Busani,\textit{$^{a}$} Lorenzo Gialli, \textit{$^{a}$} Rosario Roberto Riso,\textit{$^{b}$} and Enrico Ronca$^{\ast}$\textit{$^{a}$}} \\

\includegraphics{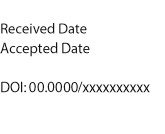} & \noindent\normalsize{Strong light–matter coupling in optical cavities promises to reshape molecular reactivity and properties in a non-invasive and tunable way, yet cavity experiments remain notoriously difficult to interpret and reproduce, highlighting the need for accurate theoretical predictions. Significant progress has been made through ab initio cavity quantum electrodynamics (QED) methods, but these generally assume ideal, lossless cavity fields, an approximation commonly justified by the assumption that cavity losses are negligible and by the need to keep the methods computationally tractable. Here we show that this approximation is far from innocent: incorporating cavity dissipation through the experimentally accessible quality factor Q reduces the Rabi splitting of p-nitroaniline by more than 10\% and red-shifts the minimum polaritonic gap by 0.24 eV, even when the system remains firmly in the strong-coupling regime. This implies that convergence towards the ideal-cavity limit requires quality factors substantially larger than those needed to establish strong coupling itself. Our approach, based on a damped harmonic oscillator description of the cavity field and a time-dependent unitary transformation yielding a real Hermitian Hamiltonian, is demonstrated here within the quantum electrodynamics coupled-cluster singles and doubles framework (QED-CCSD). Its general formulation allows straightforward integration into existing ab initio QED frameworks without increasing the computational cost of the underlying methods, providing a practical and efficient route to quantitatively account for realistic cavity conditions in first-principles simulations of molecular systems in cavities.} \\
\end{tabular}

 \end{@twocolumnfalse} \vspace{0.6cm}

  ]

\renewcommand*\rmdefault{bch}\normalfont\upshape
\rmfamily
\section*{}
\vspace{-1cm}


\footnotetext{\textit{$^{a}$~ Dipartimento di Chimica, Biologia e Biotecnologie, Via Elce di Sotto 8, 06123, Perugia, Italy; E-mail: giovanna.bruno@unipg.it; enrico.ronca@unipg.it}}
\footnotetext{\textit{$^{b}$~Department of Chemistry, Norwegian University of Science and Technology, Trondheim,
7491, Norway. }}

\footnotetext{\dag~Supplementary Information available. See DOI: 00.0000/00000000.}

\footnotetext{\dag~Additional footnotes to the title and authors can be included \textit{e.g.}\ `Present address:' or `These authors contributed equally to this work' as above using the symbols: \ddag, \textsection, and \P. Please place the appropriate symbol next to the author's name and include a \texttt{\textbackslash footnotetext} entry in the the correct place in the list.}


\section{Introduction}
In recent years, Cavity Quantum Electrodynamics (Cavity-QED) \cite{Hinds_1999,Haroche_Nobel_2013,Ruggenthaler_2023}  has become progressively central to quantum optics, material science and physical-chemistry. This renewed interest stems from the observation that material properties and reactivity can be modified even in “dark cavities”, \textit{i.e.} without external driving \cite{Wang_2014,Schlawin_2022,Appugliese_2022,Jarc_2024,Dang_2025,Keren_2026,Montanaro_2026,Graziotto_2026,Ebbesen_2012,Shalabney_2015,Ebbesen_2016,Ebbesen_2019,Long_2015,Lather_2019,Hirai_2021,Mony_2021,Peters_19,Lather_2021_enz,Vergauwe_2019}, with the advantage of a potentially negligible energetic cost \cite{Garcia_Vidal_2021}. Such effects, particularly those concerning chemical reactivity and kinetics, have been attributed to the formation of polaritonic states arising from strong coupling between matter and the cavity vacuum field \cite{Hirai_2023}. Indeed, these states are hybrid light-matter quantum states possessing distinct properties that cannot be simply reduced to those of their bare molecular and photonic constituents. Support for this interpretation comes from the observation of Rabi splitting in cavity spectra for cavities not subject to external driving \cite{Ebbesen_2012,Bienfait_2016,Ebbesen_2019}, together with extensive theoretical studies predicting cavity-induced modifications of matter properties under the same conditions \cite{Schutz_2020,Li_2021,Schafer_2022,Rokaj_2022,Rokaj_2023,CamposGonzalezAngulo_2019,Wang_2019,Latini_2021,Lu_2024,Arwas_2023,Passetti_2023,Ashida_2020}.
 Nevertheless, the microscopic origin of these modifications remains at the center of intense debate \cite{simpkins_2021,Simpkins_2023,Thomas_2024,Pannir_2025}, especially because of the intrinsic complexity of polaritonic experiments. Indeed, they are often difficult to reproduce and highly sensitive to environmental conditions and to the specific experimental protocol \cite{Michon_2025}. \par
In this context, theoretical modeling is instrumental in disentangling the intertwined physical and chemical processes occurring in cavity experiments by selectively controlling and isolating the different mechanisms that may contribute to the observed phenomena\cite{Yang_2021,Sanchez_2022,Pannir_2025,Fassioli_2025}. Beyond effective model Hamiltonians, including the Jaynes–Cummings \cite{Jaynes_Cummings_1963}, Tavis–Cummings \cite{Tavis_Cummings_1968} and Hopfield \cite{Hopfield_1958} models, a central challenge in modern Cavity-QED theory is the development of quantitative and predictive frameworks for realistic cavity systems \cite{Ruggenthaler_2023}. Significant progress toward this goal has been achieved through ab-initio approaches based on the Pauli–Fierz Hamiltonian, including QED-adapted density functional theory \cite{Ruggenthaler_2015, Ruggenthaler_2014, Fuck_2016,Flick_2015, Tokatly_2013}, Hartree–Fock (HF) \cite{Haugland_2021,Riso_2022,Schnappinger_2023,Castagnola_2025}, coupled-cluster (CC)\cite{Haugland_2020}, Complete Active Space Self-Consistent Field (CASSCF) \cite{Vu_2024,Alessandro_2025} and perturbation theory\cite{ElMoutaoukal_2025} methods, among others. Indeed, by enabling an accurate description of the emitter's electronic structure and chemical properties under light–matter coupling, these approaches significantly expand the predictiveness of Cavity-QED simulations. \par
Despite these advances, the description of the electromagnetic field in ab-initio Cavity-QED approaches still relies on several simplifying assumptions. Among these, a common one is the ideal-cavity approximation, in which the photonic modes are described as the lossless solutions of the source-free Maxwell's equations. Thus, within this approximation the cavity is not treated as a physical electromagnetic structure with its own optical response, but is accounted for solely through the normalization of these modes within a finite quantization volume, which is typically identified with the cavity's one. Recent developments have moved beyond this idealized description by explicitly accounting for the electromagnetic response and finite reflectivity of cavity mirrors through transfer-matrix approaches \cite{Svendsen_2025}. However, introducing finite cavity losses within an ab-initio Cavity-QED framework remains considerably challenging. A rigorous treatment requires the explicit description of the electromagnetic environment, either through open-quantum-system approaches \cite{Breuer_2007,Pino_2015,Ying_2023,Lindoy_2023,Fiechter_2023,Lindoy_2024,Ke_2024} or macroscopic QED \cite{Huttner_1992,Matloob1995,Scheel1999a,Scheel1999b,Gruner1995,Hsu_2025}, where the cavity modes are coupled to a continuum of bath modes. Although both master-equation formulations, such as those based on the Lindblad equation, and non-Hermitian approaches \cite{Jager_2022,Lednev_2024,Settineri_2018,Ulusoy_2020,Panayiota_2020} have been employed to account for cavity losses in Cavity-QED, these treatments relied on effective model Hamiltonians for describing the field.
A notable alternative is provided by Wang \textit{et al.} \cite{Wang_2021}, who recently extended the QED Density Functional Theory (DFT) method to include cavity dissipation through a Fano-type QED model based on the coupling of the electronic system to a continuous photonic reservoir. However, the explicit treatment of thousands of photonic modes to represent the dissipative environment poses a significant challenge for large-scale ab-initio simulations. Another interesting strategy was recently proposed by Brorsen \cite{Brorsen_2026}, who incorporated cavity losses into the real-time QED coupled-cluster approach through a complex cavity frequency, thereby enabling the description of spectroscopic signatures of dissipation, such as the polariton linewidths.\par

The ideal-cavity approximation is also supported by the established understanding of the strong-coupling regime. Indeed, within the Jaynes–Cummings picture, this regime is achieved when the coherent light–matter interaction exceeds the decay rates of both the material excitation ($\gamma=\frac{1}{\tau_{mat}}$) and the cavity mode ($\kappa=\frac{1}{\tau_{ph}}$). Therefore, the observation of  the Rabi splitting is commonly interpreted as an indication that photon leakage represents only a minor perturbation to the hybrid light–matter states. This assumption is further reinforced in realistic Cavity-QED settings, where the light–matter interaction is intrinsically a collective phenomenon, enhanced by the large number of material degrees of freedom coupled to the confined electromagnetic field, either in molecular ensembles or extended materials. Nevertheless, real cavities always exhibit finite photon lifetimes, quantified by the quality factor $Q$ through the relation $Q=\frac{\omega_{\mathbf{k}}}{\kappa}$ and these losses can still influence the system even in the strong coupling regime. The relevance of this has been recently highlighted by Rashidi \textit{et al.} \cite{Rashidi_2025,Toppari_2025}, who demonstrated that reducing the cavity energy lifetime, passing from a distributed Bragg reflector (DBR) cavity to a waveguide with a lower $Q$-factor, suppresses the induced acceleration of a charge-transfer reaction, despite the persistence of strong coupling, as evidenced by the observation of avoided crossing in the cavity spectra. These observations naturally prompt a central question: to what extent can finite cavity losses modify the energetics of strongly coupled light–matter systems, even when the system remains firmly within the strong-coupling regime? Addressing this question requires a treatment of cavity dissipation that is sufficiently accurate to capture its effects while remaining computationally affordable for realistic light–matter systems. \par
To this end, we introduce a minimal phenomenological extension of the Pauli--Fierz framework to account for cavity dissipation. The latter is incorporated by treating the cavity field as a damped harmonic oscillator, with the damping strength determined by the cavity quality factor $Q$. The approach is applicable to different resonator architectures, regardless of the device type. The method is integrated into the QED coupled-cluster framework implemented in the program package \texttt{eT} \cite{Folkestad_2020,Folkestad_2026}  without increasing the computational complexity and cost of the original approach. The application to the $S_0$-$S_1$ transition of p-nitroaniline (PNA) shows that reducing the cavity quality factor from the ideal lossless limit ($Q\to\infty$)  leads to significant reductions of the Rabi splitting at the QED-Coupled Cluster Singles and Doubles (QED-CCSD) level. These results demonstrate that finite cavity losses can introduce substantial corrections to the energetics of polaritonic states and highlight the importance of a realistic cavity description in Cavity-QED simulations.

\section{Theory and Methods}
\subsection{Q-factor based model for lossy cavities in the strong-coupling regime}
The starting point of our approach is the Pauli–Fierz Hamiltonian describing the energy of the coupled light–matter system at the quantum level. We work in the Coulomb gauge, where the quantized electromagnetic field is described by a transverse vector potential $\hat{\mathbf A}(\mathbf r)$, satisfying $\nabla\cdot\hat{\mathbf A}(\mathbf r)=0$. The resulting Hamiltonian reads:

\begin{equation}
\begin{aligned}
\hat{H}_{\mathrm{PF}} ={}&
\sum_{i=1}^{N_e}
\frac{1}{2}
\left(
\hat{\mathbf{p}}_i-\hat{\mathbf{A}}(\mathbf{r}_i)
\right)^2
-\sum_{i=1}^{N_e}\sum_{A=1}^{N_n}
\frac{Z_A}{|\mathbf{r}_i-\mathbf{R}_A|}
\\
&+\sum_{i<j}^{N_e}
\frac{1}{|\mathbf{r}_i-\mathbf{r}_j|}
+\sum_{A<B}^{N_n}
\frac{Z_A Z_B}{|\mathbf{R}_A-\mathbf{R}_B|}
\\
&+\sum_{\mathbf{k},\alpha}
\omega_{\mathbf{k}}
\left(
\hat{b}_{\mathbf{k},\alpha}^{\dagger}
\hat{b}_{\mathbf{k},\alpha}
+\frac{1}{2}
\right)
\end{aligned}
\label{eq.H_PF}
\end{equation}

The coupling between the electronic and photonic degrees of freedom is introduced through the minimal-coupling substitution $\hat{\mathbf p}_i\rightarrow\hat{\mathbf p}_i-\hat{\mathbf A}(\mathbf r_i)$. Expanding the corresponding kinetic-energy term, two distinct field-dependent contributions emerge: the paramagnetic term, which describes the direct interaction between the electronic motion and the electromagnetic field, and the diamagnetic term, which accounts for an additional contribution to the system energy beyond that of the free electromagnetic field, namely the last term of Equation \ref{eq.H_PF}. 
As anticipated, the cavity is not explicitly included in the standard Pauli-Fierz description, except through the field quantization volume $V$. As a result, the electromagnetic field is effectively treated as a free field, governed by the standard wave equation.  The corresponding quantized vector potential can therefore be expanded in terms of plane-wave modes as

\begin{equation}
\mathbf{A}(\mathbf{r},t) = \sum_{\mathbf{k},\alpha}\frac{\lambda}{\sqrt{\omega_{\mathbf{k}}}}\Big(\boldsymbol{\epsilon}_{\mathbf{k},\alpha}\hat{b}_{\mathbf{k},\alpha}e^{i\mathbf{k}\cdot\mathbf{r}}e^{-i\omega_{\mathbf{k}}t}+\boldsymbol{\epsilon}_{\mathbf{k},\alpha}^{*}\hat{b}_{\mathbf{k},\alpha}^{\dagger}e^{-i\mathbf{k}\cdot\mathbf{r}}e^{+i\omega_{\mathbf{k}}t}\Big)
    \label{eq.A_std}
\end{equation}

where $\hat{b}_{\mathbf{k},\alpha}$,  $\hat{b}^{\dagger}_{\mathbf{k},\alpha}$ are the photonic ladder operators per mode $\mathbf{k}$ and polarization $\alpha$, satisfying the commutation rule $[\hat{b}_{\mathbf{k},\alpha},\hat{b}^{\dagger}_{\mathbf{k'},\alpha'}]=\delta_{\mathbf{kk'}}\delta_{{\alpha \alpha'}}$, and $\lambda = \sqrt{\frac{1}{\epsilon_{0}\epsilon_{r}V}}$ is the field coupling strength.

To account for finite cavity losses within the Pauli–Fierz framework, we go beyond this standard description by introducing a phenomenological model in which the electromagnetic field obeys a damped wave equation. More specifically, we start from the classical damped D'Alambert equation for the vector potential

\begin{equation}
 \nabla^{2}\mathbf{A}(\mathbf{r},t)=\frac{1}{c^2}\partial_{tt}\mathbf{A}(\mathbf{r},t) + b\partial_{t}\mathbf{A}(\mathbf{r},t) 
    \label{eq.1}
\end{equation}

where the damping factor $b$ is related to the cavity resonance frequency and $Q$-factor by the relation $c^{2}b=\frac{\omega_{\mathbf{k}}}{Q}$, namely the photon decay rate $\kappa$. This equation admits a general solution of the form

\begin{equation}
\begin{aligned}  
\mathbf{A}(\mathbf{r},t) =& \sum_{\mathbf{k},\alpha}\boldsymbol{\epsilon}_{\mathbf{k},\alpha}\Big(A_{\mathbf{k},\alpha}e^{i\mathbf{k}\cdot\mathbf{r}}e^{-\frac{\omega_{\mathbf{k}}}{2Q}t-i\omega_{\mathbf{k}}t\sqrt{1-\frac{1}{4Q^2}}}+ \\ &A_{\mathbf{k},\alpha}^{*}e^{-i\mathbf{k}\cdot\mathbf{r}}e^{-\frac{\omega_{\mathbf{k}}}{2Q}t+i\omega_{\mathbf{k}}t\sqrt{1-\frac{1}{4Q^2}}}\Big)
\end{aligned}
    \label{eq.2}
\end{equation}

where the field is assumed to be linearly polarized, \textit{e.g.} $\boldsymbol{\epsilon}_{\mathbf{k},\alpha}=\boldsymbol{\epsilon}^{*}_{\mathbf{k},\alpha}$.

Since observing polaritonic phenomena requires the light–matter coupling to exceed both cavity dissipation and matter relaxation rates, cavities operating in the strong-coupling regime are typically designed with high quality factors. In these conditions it is reasonable adopting the so-called \textit{very weak damping} approximation ($\frac{1}{Q^2} \ll 1$), which neglects all terms of order $Q^{-n}, n \geq 2$. This corresponds to disregarding the damping-induced shift of the resonance frequency and retaining only the decay of the field's amplitude with rate $\frac{\kappa}{2}=\frac{\omega_{\mathbf{k}}}{2Q}$ for finite quality factors 

\begin{equation}
\begin{aligned}
\mathbf{A}(\mathbf{r},t) = &\sum_{\mathbf{k},\alpha}\boldsymbol{\epsilon}_{\mathbf{k},\alpha}\Big(A_{\mathbf{k},\alpha}e^{i\mathbf{k}\cdot\mathbf{r}}e^{-i\omega_{\mathbf{k}}t}+A_{\mathbf{k},\alpha}^{*}e^{-i\mathbf{k}\cdot\mathbf{r}}e^{i\omega_{\mathbf{k}}t}\Big)e^{-\frac{\omega_{\mathbf{k}}}{2Q}t}=\\&\sum_{\mathbf{k},\alpha}\mathbf{A}^{\mathbf{k},\alpha}_{\mathrm{HO}}(\mathbf{r},t)e^{-\frac{\omega_{\mathbf{k}}}{2Q}t}
\end{aligned}
 \label{eq.3}
\end{equation}
where $\mathbf{A}^{\mathbf{k},\alpha}_{\mathrm{HO}}=\boldsymbol{\epsilon}_{\mathbf{k},\alpha}\Big(A_{\mathbf{k},\alpha}e^{i\mathbf{k}\cdot\mathbf{r}}e^{-i\omega_{\mathbf{k}}t}+A_{\mathbf{k},\alpha}^{*}e^{-i\mathbf{k}\cdot\mathbf{r}}e^{i\omega_{\mathbf{k}}t}\Big)$ is the harmonic part of the vector potential.\\

Despite the damping, under the \textit{very weak damping} approximation, the first-order Maclaurin expansion of the electromagnetic field energy around $t=0$ recovers the time-independent harmonic energy (see Section S1 of the ESI for details). This short-time limit defines the ideal-cavity reference of the strongly coupled light--matter system, in which the polaritonic state is assumed to be established at $t=0$. Finite cavity losses are then incorporated as perturbative corrections to the polaritonic energies by expanding the vector potential in Equation \ref{eq.3} to first order around this reference point 

\begin{equation}
\begin{split}
\mathbf{A}(\mathbf{r},t) &\simeq \sum_{\mathbf{k},\alpha}\boldsymbol{\epsilon}_{\mathbf{k},\alpha}\Big[A_{\mathbf{k},\alpha}e^{i\mathbf{k}\cdot \mathbf{r}}\Big(1-i\omega_{\mathbf{k}}t-\frac{\omega_{\mathbf{k}}}{2Q}t\Big)+\\&A^{*}_{\mathbf{k},\alpha}e^{-i\mathbf{k}\cdot \mathbf{r}}\Big(1+i\omega_{\mathbf{k}}t-\frac{\omega_{\mathbf{k}}}{2Q}t\Big)\Big]=\\&=\sum_{\mathbf{k},\alpha}\boldsymbol{\epsilon}_{\mathbf{k},\alpha}\Big[A^{\mathbf{k},\alpha}_{\mathrm{HO},t=0}(\mathbf{r})\Big(1-\frac{\omega_{\mathbf{k}}}{2Q}t\Big)-E^{\mathbf{k},\alpha}_{\mathrm{HO},t=0}(\mathbf{r})t\Big]
\end{split}
\label{eq.4}
\end{equation}

where $A^{\mathbf{k},\alpha}_{\mathrm{HO},t=0}(\mathbf{r})=A_{\mathbf{k},\alpha}e^{i\mathbf{k}\cdot \mathbf{r}}+A^{*}_{\mathbf{k},\alpha}e^{-i\mathbf{k}\cdot \mathbf{r}}$ and $E^{\mathbf{k},\alpha}_{\mathrm{HO},t=0}(\mathbf{r})=i\omega_{\mathbf{k}}(A_{\mathbf{k},\alpha}e^{i\mathbf{k}\cdot \mathbf{r}}-A^{*}_{\mathbf{k},\alpha}e^{-i\mathbf{k}\cdot \mathbf{r}})$ denote the initial time $t=0$ spatial profiles of the vector potential and electric field associated with the $(\mathbf{k},\alpha)$ harmonic mode, respectively. \par
By exploiting the definition of the source-free electric field in Coulomb gauge $\mathbf{E}=-\partial_t \mathbf{A}$, the damped vector potential can be recasted as 

\begin{equation}
\begin{aligned}
\mathbf{A}(\mathbf{r},t)
=&
\mathbf{A}_{\mathrm{\mathrm{HO}},t=0}(\mathbf{r})
+
t\left[
\left.\partial_t\mathbf{A}_{\mathrm{\mathrm{HO}}}(\mathbf{r},t)\right|_{t=0}
-
\frac{\omega_{\mathbf{k}}}{2Q}
\mathbf{A}_{\mathrm{\mathrm{HO}},t=0}(\mathbf{r})
\right] = \\& \mathbf{A}_{\mathrm{HO},t=0}(\mathbf{r})
+
t\left[
\left(
\partial_t-\frac{\omega_{\mathbf{k}}}{2Q}
\right)
\mathbf{A}_{\mathrm{HO}}(\mathbf{r},t)
\right]_{t=0}
\label{eq.5}
\end{aligned}
\end{equation}

Interestingly, the operator $\Big(\partial_{t}-\frac{\omega_{\mathbf{k}}}{2Q}\Big)$ exhibits a structure formally analogous to a covariant time derivative, where the dissipation rate $\frac{\omega_{\mathbf{k}}}{2Q}$ plays the role of a constant connection-like term. Indeed, it modifies the standard harmonic evolution by accounting for the attenuation of the electromagnetic mode amplitude due to finite cavity losses, an effect that is consistently reflected in the generalized momentum of the coupled light--matter system (see Equation \ref{eq.7}).

Importantly, since the electromagnetic field energy reduces to the harmonic field energy in the considered limit, the canonical quantization procedure can be retained. As a result, the damped vector potential operator can be expressed in terms of the standard  photonic creation and annihilation operators, so that Equation \ref{eq.4} becomes 

\begin{equation}
\hat{\mathbf{A}}(r,t)= \sum_{\mathbf{k},\alpha}\boldsymbol{\epsilon}_{\mathbf{k},\alpha}\Big[\hat{A}^{\mathbf{k},\alpha}_{\mathrm{HO},t=0}(\mathbf{r})\Big(1-\frac{\omega_{\mathbf{k}}}{2Q}t\Big)-\hat{E}^{\mathbf{k},\alpha}_{\mathrm{HO},t=0}(\mathbf{r})t\Big]
    \label{eq.6}
\end{equation}

with 
\begin{subequations}
    \begin{equation}
        \hat{A}^{\mathbf{k},\alpha}_{\mathrm{HO},t=0}(\mathbf{r})=\frac{\lambda }{\sqrt{2\omega_{\mathbf{k}}}}\Big(\hat{b}_{\mathbf{k},\alpha}e^{i\mathbf{k}\cdot \mathbf{r}}+\hat{b}^{\dagger}_{\mathbf{k},\alpha}e^{-i\mathbf{k}\cdot \mathbf{r}}\Big)
\label{eq.6_A_main}
    \end{equation}
\begin{equation}
\hat{E}^{\mathbf{k},\alpha}_{\mathrm{HO},t=0}(\mathbf{r})=i\lambda\sqrt{\frac{\omega}{2}}\Big(\hat{b}_{\mathbf{k},\alpha}e^{i\mathbf{k}\cdot \mathbf{r}}-\hat{b}^{\dagger}_{\mathbf{k},\alpha}e^{-i\mathbf{k}\cdot \mathbf{r}}\Big) 
\label{eq.6_B_main}
\end{equation}
\end{subequations}

By restricting the electromagnetic field to a single mode and adopting the Long-Wavelength Approximation (LWA), in which the field is assumed to be spatially uniform over the molecular extent, the substitution of the damped vector potential operator (Equation \ref{eq.6}) into the Pauli–Fierz Hamiltonian yields 

\begin{equation}
\begin{aligned}
 \hat{H} = &\frac{1}{2}\sum_{i}\Big(\hat{\mathbf{p}}_{i}-\hat{\mathbf{A}}_{\mathrm{HO},t=0}(1-\frac{\omega}{2Q}t)+\hat{\mathbf{E}}_{\mathrm{HO},t=0}t\Big)^{2} -\sum_{i=1}^{N_e}\sum_{A=1}^{N_n}
\frac{Z_A}{|\mathbf{r}_i-\mathbf{R}_A|}
\\
&+\sum_{i<j}^{N_e}
\frac{1}{|\mathbf{r}_i-\mathbf{r}_j|}
+\sum_{A<B}^{N_n}
\frac{Z_A Z_B}{|\mathbf{R}_A-\mathbf{R}_B|}
\\
&+\sum_{\alpha}
\omega_{}
\left(
\hat{b}_{\alpha}^{\dagger}
\hat{b}_{\alpha}
+\frac{1}{2}
\right)
 \end{aligned}
    \label{eq.7}
\end{equation}

The new form of the Pauli–Fierz Hamiltonian provides an interesting physical interpretation of the adopted dissipative model. Indeed, by combining the \textit{very weak damping approximation} and the first-order Maclaurin expansion, the description is restricted to a regime in which dissipative effects are incorporated as small subleading corrections to the energetics of the coupled system. Indeed, within this limit, the harmonic field energy remains unchanged, while cavity losses only affect the light-matter coupling term through the damped vector potential operator. Moreover, such losses enter the Hamiltonian as a parametric correction to the electromagnetic field dynamics: they modify the temporal evolution of the field modes without altering their underlying quantum structure and, in particular, without requiring redefining the photonic creation and annihilation operators.

At this point, we further transform the Hamiltonian by applying the canonical Power–Zienau–Woolley (PZW) transformation \cite{Woolley_1980}, also commonly referred to as the length-gauge transformation, which recasts the minimal-coupling interaction in terms of the electric dipole moment and the electric field. This change of representation is particularly convenient for the implementation inside the \texttt{eT} software package. Indeed, in this framework, the inclusion of the dissipation effects results in a simple modification of the prefactors appearing in the Pauli-Fierz Hamiltonian. However, applying the standard PZW transformation in this context requires an extension to account for the explicit time dependence of the damped vector potential reported in Equation ~\ref{eq.6}. Within the LWA, the resulting time-dependent PZW transformation reads:

\begin{equation}
\begin{aligned}
    \hat{U}_{PZW} =& e^{i\hat{\mathbf{A}}(t)\cdot\hat{\mathbf{R}}}=e^{i(\hat{\mathbf{A}}_{\mathrm{HO},t=0}(1-\frac{\omega}{2Q}t)-\hat{\mathbf{E}}_{\mathrm{HO},t=0}t)\cdot \hat{\mathbf{R}}} \\=& e^{-i(\hat{\mathbf{A}}_{\mathrm{HO},t=0}(1-\frac{\omega}{2Q}t)-\hat{\mathbf{E}}_{\mathrm{HO},t=0}t)\cdot \sum_{j}\hat{\mathbf{r}}_{j}}
    \end{aligned}
    \label{eq.8}
\end{equation}

Here, under the Born-Oppenheimer approximation the total dipole moment operator $\hat{\mathbf{R}}$ reduces to the total electronic dipole moment $\hat{\mathbf{R}}=-\sum_{j}\hat{r}_{j}$  since the nuclear contribution is constant with respect to the electronic coordinates.

In this context, the time variable does not represent an additional dynamical degree of freedom. It only enters the description parametrically through the multiplicative time-dependent coefficients in front of the photonic operators. Since we do not consider the time evolution of the system wavefunction, the Hamiltonian is transformed pointwise in time as $\hat{H}'(t)=\hat{U}(t)\hat{H}\hat{U}^{\dagger}(t)$ without the term $i(\partial_t\hat{U})\hat{U}^{\dagger}$ arising from the transformation of the time-dependent Schr\"{o}dinger equation. Thus, by applying $\hat{U}_{PZW}$ to the Pauli-Fierz Hamiltonian (see Section S2 of the ESI) the latter transforms as

\begin{equation}
\begin{aligned}
    \hat{H}(t) = &\frac{1}{2}\sum_{i}\hat{\mathbf{p}}^{2}_{i}  -\Big(1-\frac{\omega}{2Q}t\Big)\hat{\mathbf{E}}_{\mathrm{HO},t=0}\cdot\sum_{j}\hat{\mathbf{r}}_{j} -\omega^{2}t\hat{\mathbf{A}}_{\mathrm{HO},t=0}\cdot\sum_{j}\hat{\mathbf{r}}_{j} +\\&\frac{\lambda^2}{2}\Big((1-\frac{\omega}{2Q}t)^2+\omega^2 t^2\Big)\sum_{\alpha,i,j}\boldsymbol{\epsilon}_{\alpha}^2 \cdot \hat{\mathbf{r}}_{i}\hat{\mathbf{r}}_{j} + \sum_{\alpha}\omega(\hat{b}^{\dagger}_{\alpha}\hat{b}_{\alpha}+\frac{1}{2})
    \end{aligned}
    \label{eq.9}
\end{equation}

Crucially, the combined limit $Q\rightarrow\infty$ and $t\rightarrow0$ of this Hamiltonian is well-defined and corresponds to the standard time-independent Pauli--Fierz Hamiltonian in length gauge. Furthermore, its structure directly follows from the first-order expansion of the damped vector potential, which is expressed in terms of both the harmonic vector potential and its time derivative, namely the harmonic electric field (Equations \ref{eq.4} and \ref{eq.5}). Thus, after the PZW transformation, these two contributions give rise to two distinct bilinear light--matter interaction terms. The electric-dipole coupling retains the conventional length-gauge form, although its strength is renormalized by the prefactor $\left(1-\frac{\omega}{2Q}t\right)$ depending on the damping, whereas an additional dipolar coupling involving the harmonic vector potential appears from the first-order expansion, with a coefficient proportional to $\omega^{2} t$. 
At this stage, it is convenient to recast the Hamiltonian into an explicitly real form, so as to preserve the standard Pauli-Fierz Hamiltonian form implemented in the \texttt{eT} software package but accounting for cavity dissipation. In the conventional Pauli--Fierz Hamiltonian, this is readily achieved through a constant phase rotation of the photonic creation and annihilation operators, which renders the electric-dipole interaction explicitly real ($\hat{U}_{\mathbf{k},\alpha}=e^{-i\frac{\pi}{2}\hat{n}_{\mathbf{k},\alpha}} /  \hat{U}_{\mathbf{k},\alpha}\hat{b}_{\mathbf{k},\alpha}\hat{U}^{\dagger}_{\mathbf{k},\alpha} \rightarrow -i\hat{b}_{\mathbf{k},\alpha} \land \hat{U}_{\mathbf{k},\alpha}\hat{b}^{\dagger}_{\mathbf{k},\alpha}\hat{U}^{\dagger}_{\mathbf{k},\alpha} \rightarrow +i\hat{b}^{\dagger}_{\mathbf{k},\alpha}$). However, in this case the additional coupling associated with the vector potential precludes such a straightforward procedure. Indeed, the same transformation would make this contribution purely imaginary. To overcome this issue, the two bilinear terms are first combined into a single one with time-dependent complex coefficients. The resulting overall phase can then be absorbed through a suitable time-dependent unitary transformation of the photonic creation and annihilation operators (see Section S3 of the ESI for the derivation), yielding the equivalent real-valued interaction

\begin{equation}
\sqrt{1-\frac{\omega}{Q}t+\big(\omega^2+\frac{\omega^2}{4Q^2}\big)t^2}\sqrt{\frac{\omega}{2}}\lambda\sum_{\alpha}\big(\hat{b}_{\alpha}+\hat{b}^{\dagger}_{\alpha}\big)\boldsymbol{\epsilon}_{\alpha}\cdot\sum_{j}\hat{\mathbf{r}}_{j}
    \label{eq.10}
\end{equation}

 This equation has the same form as the standard real-valued bilinear term of the Pauli--Fierz Hamiltonian in length gauge, except for the time-dependent prefactor $\sqrt{1-\frac{\omega}{Q}t+\big(\omega^2+\frac{\omega^2}{4Q^2}\big)t^2}$. 
Consistently with the first-order treatment adopted throughout this work, the square root is expanded in a Maclaurin series and truncated after the linear term in $t$

\begin{equation}
\big(1-\frac{\omega}{2Q}t\big)\sqrt{\frac{\omega}{2}}\lambda\sum_{\alpha}\big(\hat{b}_{\alpha}+\hat{b}^{\dagger}_{\alpha}\big)\boldsymbol{\epsilon}_{\alpha}\cdot\sum_{j}\hat{\mathbf{r}}_{j}
    \label{eq.11}
\end{equation}

The same first-order expansion is also applied to the dipole self-energy term, leading to the final form of the dissipative Pauli--Fierz Hamiltonian

\begin{equation}
\begin{aligned}
\hat{H}(t) = &  \sum_{\alpha,j}\Big(\frac{\omega}{2Q}t-1\Big)\sqrt{\frac{\omega}{2}}\lambda(\hat{b}_{\alpha}+\hat{b}^{\dagger}_{\alpha})\boldsymbol{\epsilon}_{\alpha}\cdot\hat{\mathbf{r}}_{j} + \\&\sum_{\alpha,i,j}\frac{\lambda^2}{2}\Big(1-\frac{\omega}{Q}t\Big)\boldsymbol{\epsilon}_{\alpha}^2 \cdot \hat{\mathbf{r}}_{i}\hat{\mathbf{r}}_{j} + \hat{H}_{Field} + \hat{H}_{Mol}
\end{aligned}
    \label{eq.12}
\end{equation}

where 

\begin{equation}
\hat{H}_{Field}=\sum_{\alpha}\omega_{\alpha}\Big(\hat{b}^{\dagger}_{\alpha}\hat{b}_{\alpha}+\frac{1}{2}\Big)
\end{equation}

and 

 \begin{equation}    
 \hat{H}_{Mol}=\frac{1}{2}\sum_{i}\hat{\mathbf{p}}^{2}_{i}-\sum_{i=1}^{N_e}\sum_{A=1}^{N_n}
\frac{Z_A}{|\mathbf{r}_i-\mathbf{R}_A|}
+\sum_{i<j}^{N_e}
\frac{1}{|\mathbf{r}_i-\mathbf{r}_j|}
+\sum_{A<B}^{N_n}
\frac{Z_A Z_B}{|\mathbf{R}_A-\mathbf{R}_B|}
 \end{equation}

In line with our parametric treatment of time, we finally consider the time-averaged Hamiltonian, yielding an effective time-independent form that can be straightforwardly integrated into the  currently available implementations inside the \texttt{eT} software package. More specifically, the time-averaged Hamiltonian reads

\begin{equation}
\begin{aligned}
\frac{1}{\tau} \int_{0}^{\tau}H(t)dt= &\sum_{\alpha,j}\Big(\frac{\omega}{4Q}\tau-1\Big)\sqrt{\frac{\omega}{2}}\lambda(\hat{b}_{\alpha}+\hat{b}^{\dagger}_{\alpha})\boldsymbol{\epsilon}_{\alpha}\cdot\hat{\mathbf{r}}_{j} +\\&\sum_{\alpha,i,j}\frac{\lambda^2}{2}\Big(1-\frac{\omega}{2Q}\tau\Big)\boldsymbol{\epsilon}_{\alpha}^2 \cdot \hat{\mathbf{r}}_{i}\hat{\mathbf{r}}_{j}+\hat{H}_{Field} +\hat{H}_{Mol}
\end{aligned}
    \label{eq.13}
\end{equation}

where $\tau$ denotes the averaging time. Since the cavity energy decays over the photon lifetime $\tau_{ph}=\frac{Q}{\omega}$ and the present description relies on a first-order expansion around $t=0$ (see Section S1 of the ESI, Equations S7 and S8), the averaging time is restricted to $\tau\leq\tau_{ph}$.

Notably, this formulation is particularly advantageous from a computational standpoint. Indeed, the cavity dissipation simply rescales the prefactors multiplying the bilinear light-matter interaction and dipole self-energy terms. Therefore, the implementation within the existing QED-extended ab-initio methodologies only requires a modification of the corresponding integrals, without altering the computational scaling of the methods.

\section{Results and Discussion}

\subsection{Polaritonic dispersion of p-nitroaniline}

\begin{figure}[H]
    \centering
    \includegraphics[width=\columnwidth]{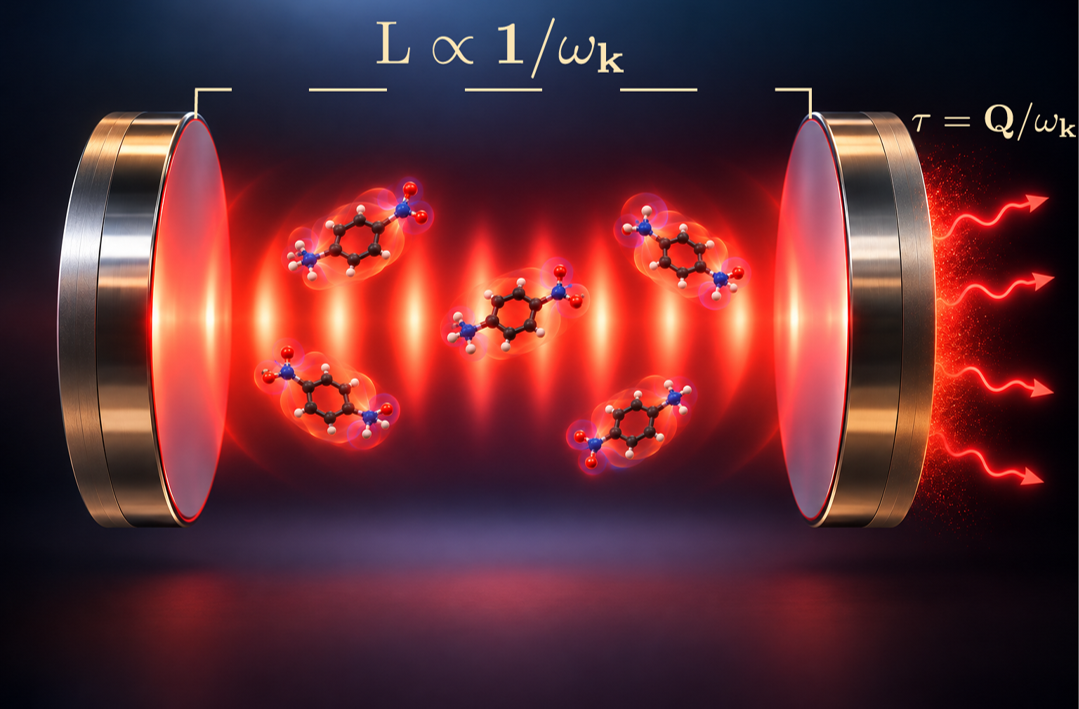}
    \caption{Representation of PNA embedded in a lossy Fabry-Pérot cavity having a photon lifetime $\tau=Q/\omega_{\mathbf{k}}$, where $Q$ is the cavity quality factor and $\omega_{\mathbf{k}}$ the resonance frequency. The cavity axis is oriented along the $z$-axis, with a length inversely proportional to the resonance frequency, $L \propto 1/\omega_{\mathbf{k}}$, while the cavity field is polarized along the $x$-axis.
 }
    \label{fig:Fig1}
\end{figure}

To quantify the effect of cavity dissipation on light--matter strong coupling, we investigate the polaritonic states of $p$-nitroaniline (PNA) (Fig. \ref{fig:Fig1}), using the Rabi splitting as a measure of the coupling strength. 

The push–pull character of PNA gives rise to a high oscillator strength for the first charge-transfer excitation, making it a suitable model system for testing our approach. In addition, PNA was previously investigated within the QED-CC framework \cite{Haugland_2020} providing a well-established reference for the present study. The molecular geometry is taken from Ref. \citenum{Haugland_2020}, where it was optimized at the DFT/B3LYP/6-31+G* level. We then employ the QED-CC framework developed in that work to compute the polaritonic state dispersion of PNA with and without cavity dissipation. Specifically, QED-CCSD calculations are performed using the cc-pVDZ basis set within the long-wavelength approximation, considering a single cavity mode polarized along the $x$-axis. The light--matter coupling is likewise set to $\lambda=0.05$, as in Ref. \citenum{Haugland_2020}. This coupling value is totally unrealizable in a real experimental device for a single emitter. However, the scope of this study is providing a relative quantification of the dissipation effects more then obtaining results directly comparable with an actual experiment. This aspect will be eventually investigated in future works. The cavity frequency is scanned over the UV--visible range of approximately $3$--$8$ eV, and the lowest ten excited states of the coupled PNA--cavity system are computed at each frequency to ensure that the relevant polaritonic states are captured.

The polaritonic state dispersion is first evaluated in the absence of cavity dissipation, corresponding to the ideal limit $Q\rightarrow\infty$, and subsequently in the presence of cavity losses using the dissipative Hamiltonian derived in this work (Equation \ref{eq.13}). For the latter case, we consider a cavity with a quality factor of $Q=60$ over a time interval $\tau=5.01$ fs, which corresponds exactly to the cavity energy lifetime $Q/\omega$ associated with $\omega=7.89$ eV. This value is chosen as a representative quality factor for experimentally realistic optical cavities used to investigate molecular strong coupling. In particular, metallic Fabry--Pérot cavities with $Q$ factors of the order of $10$--$100$ have been employed to achieve strong coupling with molecular transitions, where the finite cavity losses can be compensated by the large light–matter coupling strengths\cite{George_2015, Nagarajan_2021,Tao_2025}. Thus, $Q=60$ provides a realistic intermediate regime in which cavity dissipation is appreciable while strong coupling can still be maintained, which allows us to assess the effect of dissipation on the polaritonic-state structure and the Rabi splitting.
   
\begin{figure*}[t]
    \centering
    \includegraphics[width=
    \textwidth]{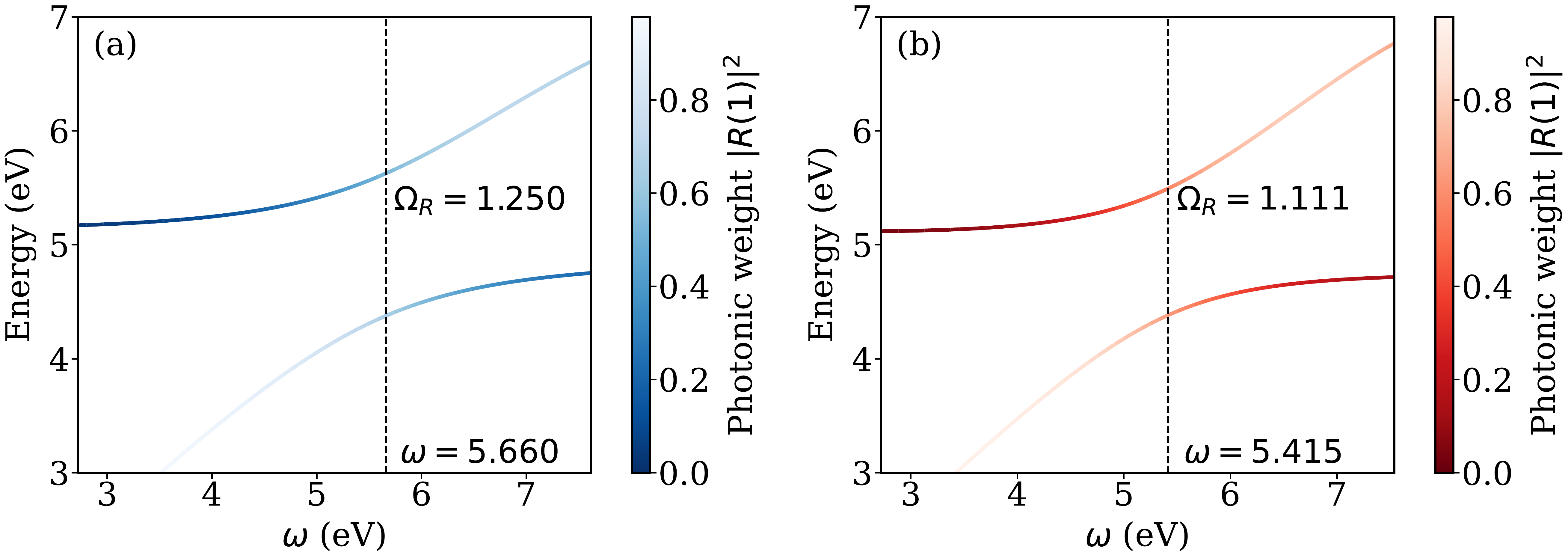}
        \caption{Polaritonic branches of PNA in the ideal cavity limit, $Q \rightarrow \infty$ (Panel a), and for $Q=60$ (Panel b). In both panels, the Rabi splitting $\Omega_R$ and the corresponding cavity frequency $\omega$ are reported in eV. The time interval in panel b is set at $\tau=5.01$ fs corresponding to the cavity lifetime for $Q=60$ and $\omega=7.89$ eV, which is the highest frequency considered in this analysis.}
        \label{fig:Fig2}
  \end{figure*}

Interestingly, and in contrast with the common assumption that cavity losses can be safely neglected in the strong-coupling regime, their inclusion has a surprisingly large impact on the polaritonic spectrum. Indeed, even for a relatively high and experimentally relevant quality factor of $Q=60$, the Rabi splitting, here defined as the minimum energy separation between the upper (UP) and lower polaritonic (LP) branches, identified as the lowest-energy QED excited states with the largest photonic character, decreases from $1.250$~eV in the ideal $Q\rightarrow\infty$ limit to $1.111$~eV at $Q=60$, corresponding to a reduction of $11.12\%$ (Figure \ref{fig:Fig2}). Such a substantial reduction, observed for a value of $Q$ compatible with the experimental observation of strong coupling, highlights the significant role that cavity dissipation can play in shaping the polaritonic spectrum. More importantly, these results suggest that neglecting dissipative effects in theoretical simulations may lead to significant inaccuracies in the resulting predictions, even in the strong-coupling regime.

Besides, the effect of dissipation is not limited to a reduction of the minimum polaritonic gap. We also observe a significant red-shift of approximately 0.24 eV in the cavity frequency at which the UP–LP separation reaches its minimum. Although a red-shift of the resonance frequency is expected for damped resonators, this effect scales as $1/Q^{2}$. This means that for a resonance frequency of 5.66 eV and $Q=60$, it amounts to less than 1 meV, more than two orders of magnitude smaller than the shift observed here. Moreover, such damping-induced renormalization of the bare cavity frequency is neglected within the very weak damping approximation, so it cannot account for our result. Rather, the shift originates from the modified structure of the effective, time-averaged Hamiltonian, in which both the light–matter interaction and dipole self-energy terms acquire an explicit dependence on $Q$ and on the time interval $\tau$. Therefore, although cavity dissipation is initially introduced phenomenologically, it ultimately rescales terms of the Pauli–Fierz Hamiltonian, entering the ab-initio QED-CCSD calculation from the outset. As a result, the polaritonic states are computed for the dissipative Hamiltonian itself, rather than obtained by correcting the ideal-cavity solution a posteriori. Consequently, dissipation modifies not only the magnitude of the polaritonic splitting, but also the energetic redistribution among the coupled light–matter states, shifting the cavity frequency at which the minimum gap is attained.

\begin{figure}[H]
    \centering
\includegraphics[width=\columnwidth]{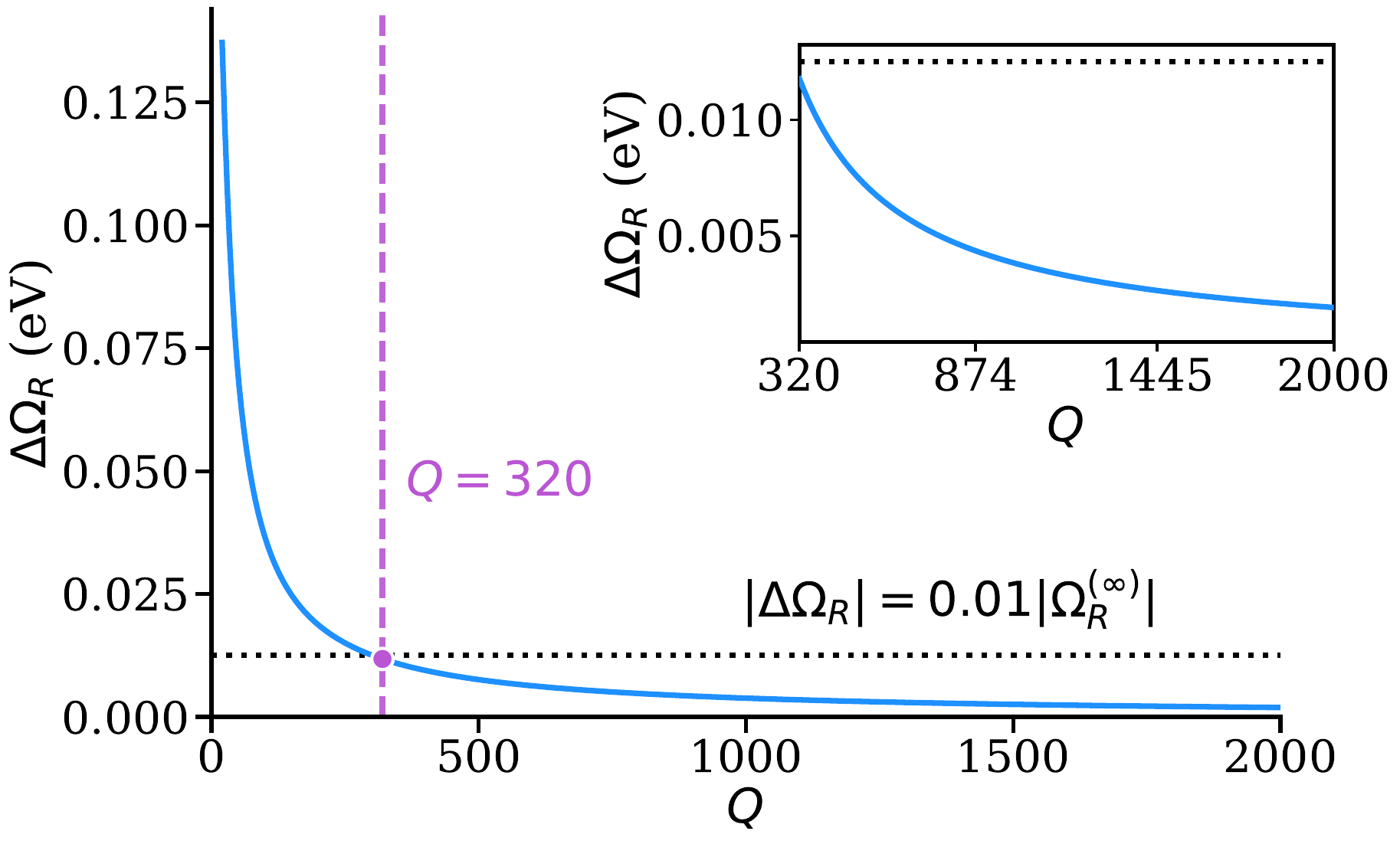}
        \caption{Convergence of the Rabi splitting towards the ideal-cavity limit as a function of the cavity quality factor $Q$. The deviation $\Delta\Omega_R=\Omega_R^{(\infty)}-\Omega_R^{(Q)}$ is evaluated at $\omega_c=5.660$ eV, corresponding to the minimum UP--LP separation in the ideal $Q\rightarrow\infty$ limit, and at $\tau=2.33$ fs corresponding to the lifetime of a cavity with $Q=20$ at this resonance frequency. The vertical purple dashed line marks $Q=320$, above which the relative deviation falls below $1\%$. The horizontal black dashed line marks the corresponding threshold. The inset shows the $Q=320-2000$ range, where the deviations remain below $0.015$ eV.}
        \label{fig:Fig3}
  \end{figure}

To further assess the persistence of cavity-dissipation effects, we investigated the convergence of the Rabi splitting towards the ideal-cavity limit as a function of the quality factor $Q$, as shown in Fig.~\ref{fig:Fig3}. Here, the UP--LP energy separation is evaluated at the cavity frequency corresponding to the minimum gap in the ideal $Q\rightarrow\infty$ case, $\omega_c=5.66$ eV. The resulting deviation from the ideal Rabi splitting, $\Delta\Omega_R=\Omega_R^{(\infty)}-\Omega_R^{(Q)}$, remains appreciable over a broad range of quality factors. The convergence analysis is performed using a fixed time interval $\tau=2.33$ fs, which is the photon lifetime in a cavity with resonance frequency $\omega=5.66$ eV and quality factor $Q=20$. This provides a conservative time window for the entire range of $Q$ considered. Indeed, since $Q/\omega$ increases with $Q$ at fixed $\omega$, this interval remains within the corresponding photon lifetime for all $Q\geq20$.

At $Q=320$, the relative deviation from the ideal Rabi splitting falls below $1$\%, making the effect of cavity dissipation physically negligible within the present convergence analysis. Notably, this value is more than five times larger than the $Q=60$ considered above, highlighting that a substantial increase in the quality factor is required to approach the ideal-cavity limit at the resonance frequency of $5.66$ eV. However, only at significantly larger $Q$-factors, around 900, does the deviation fall well below chemical accuracy. From this value onwards, it remains below $5$~meV and can therefore be considered numerically negligible within the QED-CCSD treatment.
 
\begin{figure}[H]
    \centering
        \includegraphics[width=0.46\textwidth]{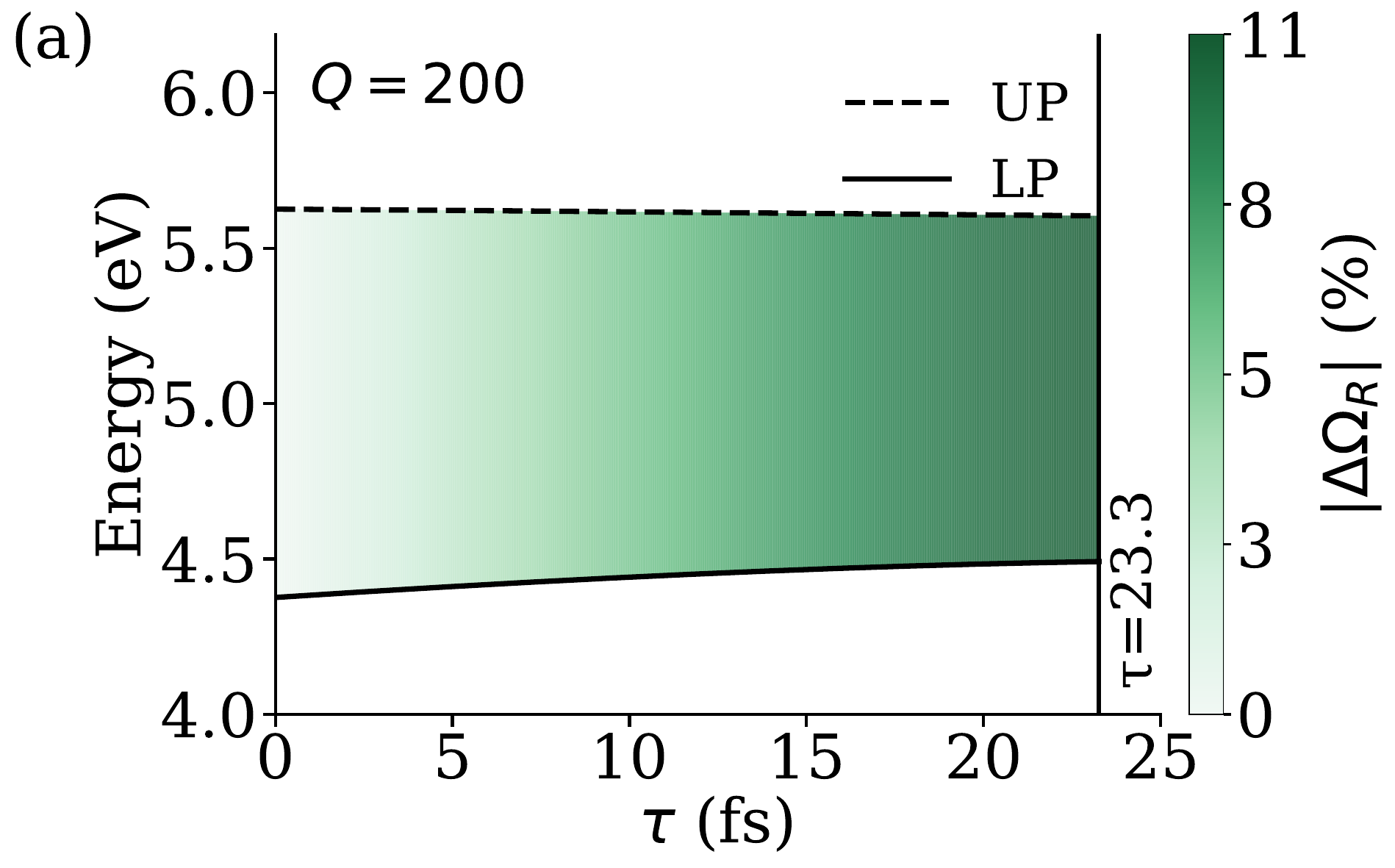}       
        \includegraphics[width=0.46\textwidth]{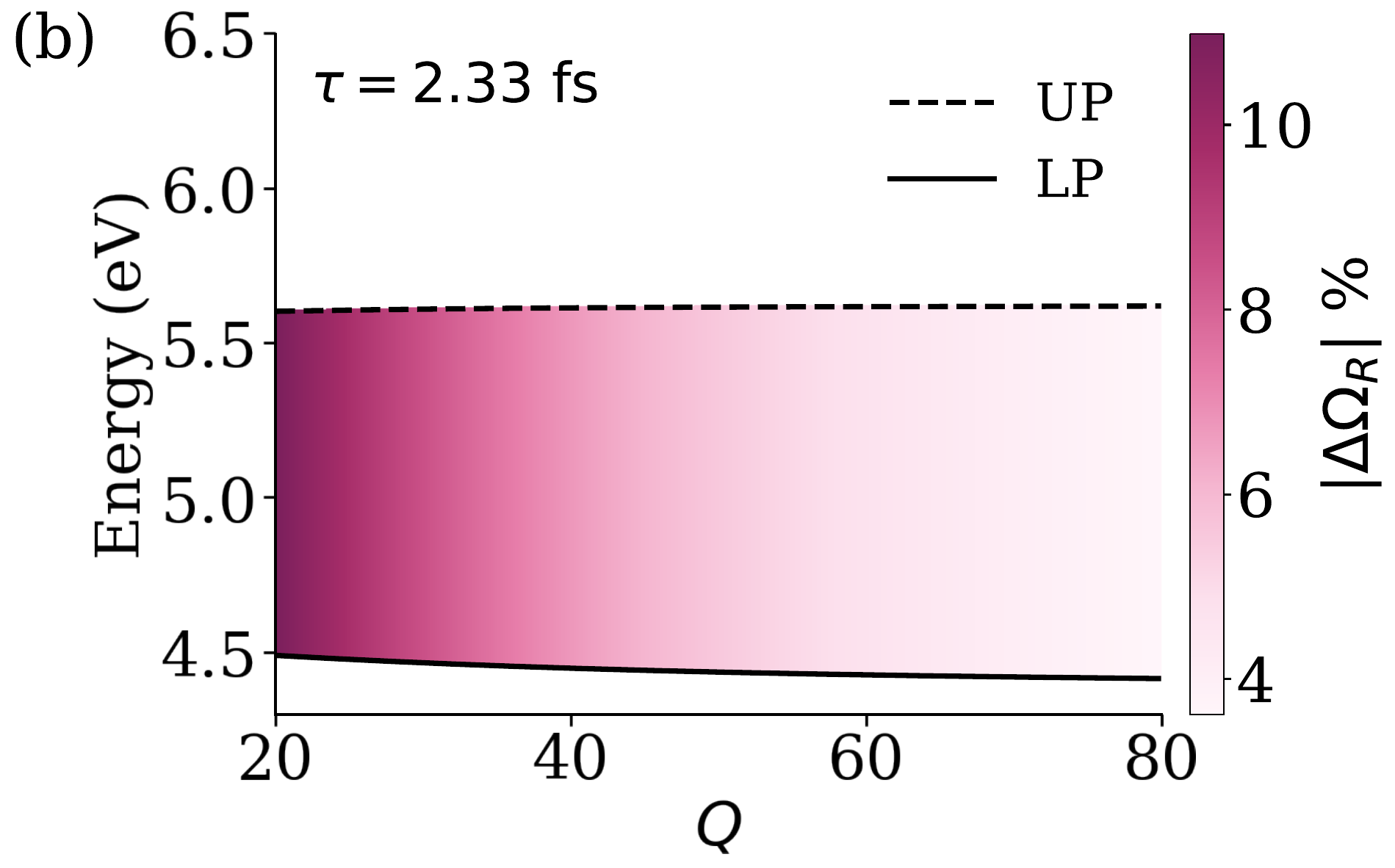}  
    \caption{Effect of cavity dissipation on the polaritonic spectrum of PNA at the reference cavity frequency of 5.66 eV for the ideal cavity framework. (a) Evolution of the upper (UP) and lower (LP) polaritonic branches as a function of time interval $\tau$ for $Q=200$. (b) Dependence of the UP and LP branches on the quality factor $Q$ at $\tau=2.33$ fs which is the cavity lifetime associated with $Q=20$. The color scale indicates the relative percentage deviation of the Rabi splitting from the ideal-cavity value.
}
    \label{fig:Fig4}
\end{figure}

We next examine how the timescale over which cavity dissipation is evaluated affects the polaritonic spectrum (Figure \ref{fig:Fig4}). Here, $\tau$ denotes the time interval over which the dissipative effects are assessed, which can extend at most up to the photon lifetime $Q/\omega$, fixed by the cavity resonance frequency and quality factor. The effects of cavity dissipation become relevant when this timescale is comparable with the characteristic timescale of the microscopic processes of interest, such as an electronic excitation or an elementary reactive event taking place in the cavity. These processes can occur on femtosecond or even shorter timescales \cite{Zewail_2000,Remacle_2006,Fernando_2024} which are fundamentally different from the overall duration of an experimental measurement, which may last much longer. It is therefore the relation between these timescales that determines the impact of cavity dissipation on the resulting polaritonic properties.

 Panel (a) of Figure \ref{fig:Fig4} follows the evolution of the UP and LP branches with respect to $\tau$ for $Q=200$ at the reference cavity frequency of $5.66$~eV associated with the ideal cavity limit. As the time increases, the two branches progressively approach each other, leading to an increasingly pronounced reduction of the Rabi splitting. This trend is directly captured by the colormap, which shows a growing deviation from the ideal-cavity value reaching approximately $11\%$ at $\tau \approx 23.3$~fs, corresponding to the photon lifetime $\frac{Q}{\omega}$. Thus, based solely on its magnitude, a quality factor of $Q=200$ might appear sufficiently high for cavity losses to have a negligible effect. However, this analysis clearly shows that at the reference cavity frequency, the dissipative effects become sizeable as the relevant timescale approaches the photon lifetime. Indeed, a reduction of comparable magnitude is observed for substantially lower quality factors, $Q=20$--$40$, when the dissipative effects are evaluated over timescales comparable to the corresponding photon lifetime (with $\tau=\frac{Q}{\omega}=2.33$~fs for $Q=20$ (panel b)). This highlights that the significance of a given quality factor cannot be assessed independently of the timescale of the physical process being probed. In this sense, both the cavity frequency and the relevant timescale become essential parameters when determining the cavity quality required to limit dissipative effects on strong-coupling observables.

This dependence on the timescale also clarifies the relation between our results and steady-state spectroscopic observables. Within a stationary description, cavity losses enter only through the photon decay rate $\kappa=\omega/Q$: they broaden the polaritonic peaks, whose linewidths scale linearly with $\kappa$, while shifting their positions only at second order in $\kappa/g$. As recently shown by Brorsen \cite{Brorsen_2026}, for sufficiently large $Q$ these effects become very small: the pole splitting of PNA at $Q=60$ is reduced by only $0.05\%$, compared with the reduction of about $11\%$ of the polaritonic gap obtained here for the same system. Steady-state spectra are therefore well suited to establish whether strong coupling is achieved, as signalled by the avoided crossing, and their linewidths encode the polariton lifetimes. However, they do not reveal how dissipation affects the polaritonic energetics over the timescale of the processes occurring in the cavity. This aspect can be crucial for chemistry, as indicated by the experiments of Rashidi \textit{et al.} \cite{Rashidi_2025,Toppari_2025}, where the cavity-induced acceleration of a charge-transfer reaction is suppressed by a shorter cavity lifetime despite the persistence of strong coupling. The time-averaged Hamiltonian adopted here addresses precisely this aspect: rather than $\kappa$ alone, the relevant parameter is the product $\kappa\tau=\omega\tau/Q=\tau/\tau_{ph}$, namely the ratio between the timescale of the process of interest and the photon lifetime. As noted by Brorsen, our approach thus describes a different quantity, whose dependence on dissipation is set by the averaging time. Within our framework, however, this dependence is not arbitrary, but reflects the timescale of the physical event being probed. As a consequence, the relevance of cavity losses depends on the process under consideration: at fixed cavity frequency, the same $Q$-factor may have a negligible impact on fast events while significantly affecting slower ones. Moreover, dissipation does not simply attenuate the light--matter coupling, but reshapes the polaritonic energy landscape relevant to the timescale of a given event, reducing the effective separation between the UP and LP branches. Such energy shifts can alter the resonance conditions between polaritonic states and other molecular states. Consider, for instance, a reaction channel that proceeds through population transfer from the UP to a molecular state $|a\rangle$ initially close in energy to it. A downward shift of the UP energy, induced by cavity dissipation, may increase the energy gap between the UP and $|a\rangle$ and suppress this transfer; at the same time, it may bring the UP close in energy to a lower-lying state $|b\rangle$, thereby opening an alternative reaction channel. Cavity dissipation may therefore not only weaken cavity-induced effects, but also redirect them, an aspect that remains inaccessible to steady-state spectra.

\section*{Conclusions}
In this work, we have addressed the role of cavity dissipation in the light--matter strong-coupling regime within an ab-initio Cavity-QED framework. Starting from the Pauli--Fierz Hamiltonian, we account for the intrinsically non-conservative nature of the characteristic cavity mode by treating its associated electromagnetic field as a damped harmonic oscillator, with the damping determined by the experimentally accessible cavity quality factor $Q$. No photonic bath or additional photonic degrees of freedom are required, retaining the standard single-mode and single-photon descriptions usually used in Cavity-QED. Within the very weak damping approximation and using a leading-order temporal expansion, canonical quantization is preserved while dissipation enters as a parametric classical correction to the light--matter interaction and dipole self-energy terms. A time-dependent unitary transformation is also introduced to recast the resulting Hamiltonian in a real-valued form, thereby incorporating cavity losses without resorting to a non-Hermitian Hamiltonian. The resulting formulation can be readily incorporated into existing ab-initio QED methods without additional computational cost or changes in computational scaling.

The impact of cavity dissipation was assessed for the $S_0-S_1$ excitation of $p$-nitroaniline at the QED-CCSD level. Remarkably, for $Q=60$, a quality factor within the range commonly employed in experiments and sufficiently high to sustain light--matter strong coupling, we find a reduction of the Rabi splitting of approximately $11$\% with respect to the ideal cavity limit. This result, consistent with recent experimental observations \cite{Rashidi_2025}, demonstrates that cavity dissipation can produce substantial modifications of the polaritonic energy spectrum even in the strong-coupling regime. Thus, although subleading with respect to the coherent light--matter interaction, cavity dissipation is by no means negligible and can be relevant for a quantitative description of strongly coupled systems. Moreover, the convergence analysis shows that substantially larger quality factors are required to recover the ideal-cavity limit, with the deviation from the ideal Rabi splitting falling below 1\% at $Q=320$. 

Beyond correcting individual calculations, the present approach provides a practical means of assessing the suitability of a cavity for a targeted application. The dependence of relevant polaritonic properties on the cavity quality factor can be readily evaluated for a given ab-initio method and desired accuracy, without the computational overhead of an explicit treatment of the photonic environment. Importantly, the effect of a given quality factor is not independent of the cavity frequency, since $Q$ determines the associated photon lifetime $\tau=Q/\omega_{\mathbf{k}}$. The latter provides a direct timescale for assessing when cavity dissipation may become relevant to the microscopic dynamics underlying the property of interest. Overall, the approach might serve as a useful preliminary screening tool for identifying the cavity quality and corresponding energy lifetime required to achieve a desired degree of control over light--matter interactions.

Although demonstrated here within QED-CCSD, the formulation can in principle be incorporated into other ab-initio QED frameworks with the same strategy and without increasing their computational cost. This provides a flexible starting point for systematically connecting experimentally accessible cavity properties with the accuracy and physical regime targeted in first-principles simulations of strongly coupled light--matter systems.

\section*{Author contributions}
G.B.: Conceptualization, Methodology, Investigation, Formal analysis, Visualization, Supervision, Funding acquisition, Writing-original draft.
E.B.: Methodology, Software.
L.G.: Methodology.
R.R.R.: Methodology, Software, Writing-review \& editing.
E.R.: Conceptualization, Software, Investigation,  Supervision, Funding acquisition, Writing-review \& editing.

\section*{Conflicts of interest}
There are no conflicts to declare.

\section*{Data availability}

The version of the \texttt{eT} program used in this work, together with an example input for calculating the polaritonic states of PNA at a specified cavity frequency, quality factor, and time interval, is available in the online repository \url{https://doi.org/10.5281/zenodo.23064296}.

\section*{Acknowledgements}
G.B. acknowledges funding from European Union – NextGenerationEU, under the PNRR – Mission 4 “Education and Research”, Component 2 “From Research to Business”, Investment 1.2 “Funding of projects submitted by young researchers”, through the project “Development of ab-initio methods for the simulation of crystalline materials in optical cavities” (CUP J93C25000490001).

R.R.R.  acknowledges funding from the European Research Council (ERC) under the European Union’s Horizon 2020 Research and Innovation Programme (grant agreement No.~101020016).

E.R. acknowledges funding from the European Research Council (ERC) under
the European Union’s Horizon Europe Research and Innovation Programme (Grant n. ERC-StG-
2021-101040197 - QED-SPIN).

The authors acknowledge the use of \textit{ChatGPT} (OpenAI. (2026). ChatGPT [Large language model]. https://chatgpt.com) as support for generating the 3D model in Figure \ref{fig:Fig1}.

The authors are grateful to Kurt R. Brorsen, for insightful discussions that significantly helped improving our manuscript.



\balance


\providecommand*{\mcitethebibliography}{\thebibliography}
\csname @ifundefined\endcsname{endmcitethebibliography}
{\let\endmcitethebibliography\endthebibliography}{}

\clearpage
\includepdf[pages=-]{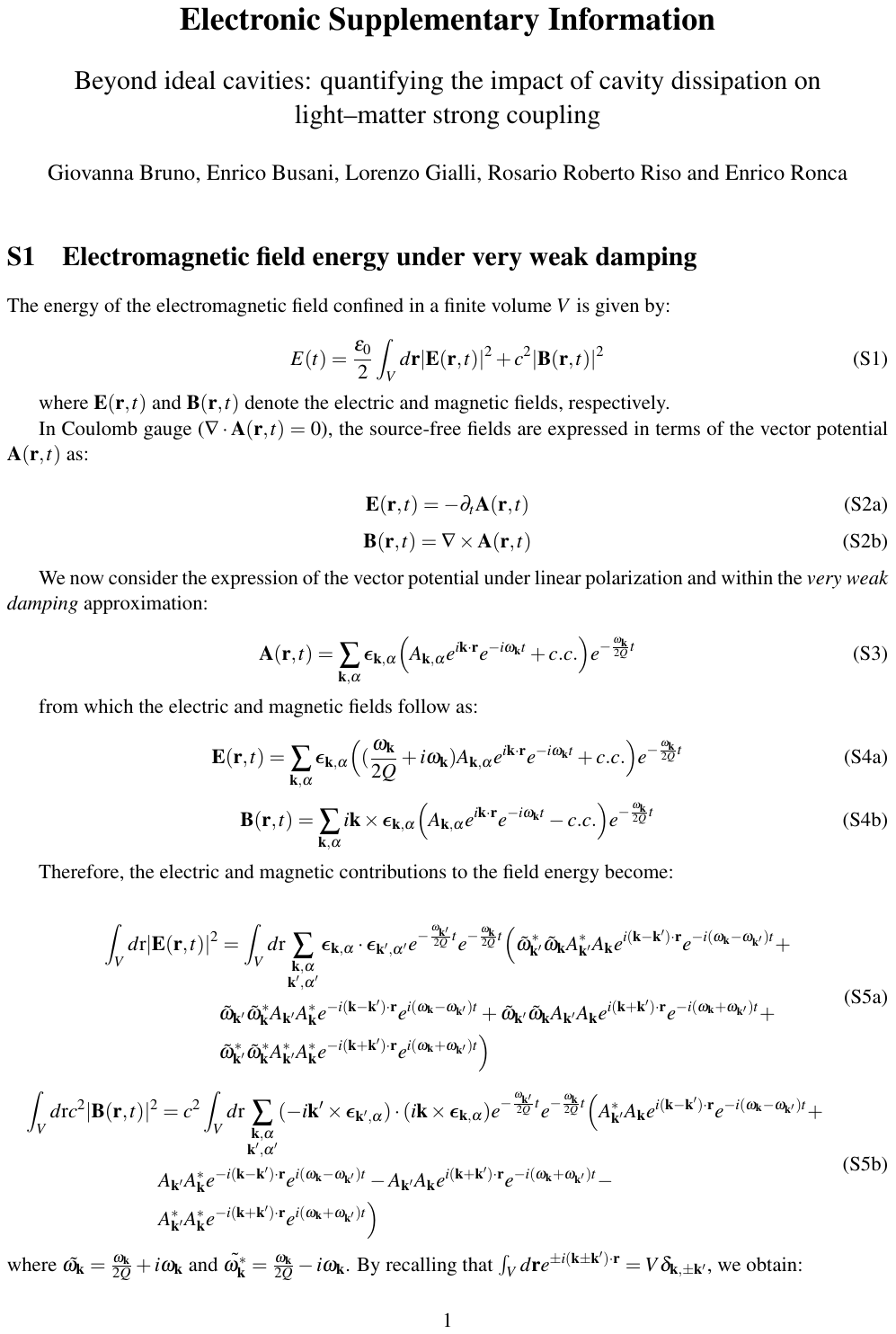}

\end{document}